\documentclass[final,5p,times,twocolumn]{elsarticle}

\usepackage{amssymb}
\usepackage{lipsum}
\usepackage{siunitx}

\usepackage{xcolor} 

\journal{Nuclear Physics A}

\begin{document}
\begin{frontmatter}


\title{Test-Beam Performance of the AstroPix Silicon Sensor for Imaging Calorimetry}


\author[pnu]{Yoonha~Hong\fnref{fn1}}\ead{yoonha@pusan.ac.kr}
\author[pnu]{Jeongsu~Bok\fnref{fn1}}\ead{jeongsu.bok@cern.ch}

\author[kangwon]{Geunpil~An}
\author[skku]{Joonsuk~Bae}
\author[skku]{Yunseul~Bae}
\author[nasa]{Regina Caputo}
\author[yonsei]{Yun~Eo}
\author[skku]{Wooseok~Ham}
\author[uos]{Woohyeon Heo}
\author[anl]{Manoj Jadhav}
\author[yonsei]{Seo~Yun~Jang}
\author[kangwon]{Jinryong~Jeong}
\author[knu]{Hyon-Suk~Jo}
\author[anl]{Sylvester~Joosten}
\author[skku]{Beomkyu~Kim}
\author[anl]{Bobae~Kim}
\author[pnu]{Chong~Kim}
\author[skku]{Dongguk~Kim}
\author[kangwon]{Minsuk~Kim}
\author[knu]{Shin~Hyung~Kim}
\author[uos]{Woojong Kim}
\author[knu]{Wonjun~Ko}
\author[knu]{Changhui~Lee}
\author[skku]{Hyungjun~Lee}
\author[yonsei]{Jason Sang Hun Lee}
\author[pnu]{Jongwon Lee}
\author[cau]{Kyeongpil Lee}
\author[knu]{Sehwook~Lee}
\author[skku]{Sangwoo~Park}
\author[pnu]{Jaehyeok~Ryu}
\author[knu]{Bogyeong Seo}
\author[anl]{Jessica~Metcalfe}
\author[knu]{Minsub Shim}
\author[knu]{Junseop~Shin}
\author[yonsei]{Hwidong~Yoo}
\author[anl]{Maria \.{Z}urek}
\author[yonsei]{Sanghoon~Lim\corref{cor}}\ead{shlim@yonsei.ac.kr}

\cortext[cor1]{Corresponding author at: Department of Physics, Yonsei University, Seoul 03722, Republic of Korea.}
\fntext[fn1]{These authors contributed equally to this work as co-first authors.}

\address[skku]{Department of Physics, Sungkyunkwan University, Suwon 16419, Republic of Korea}
\address[knu]{Department of Physics, Kyungpook National University, Daegu 41566, Republic of Korea}
\address[pnu]{Department of Physics, Pusan National University, Busan 46241, Republic of Korea}
\address[kangwon]{Department of Physics, Kangwon National University, Gangneung 25457, Republic of Korea}
\address[yonsei]{Department of Physics, Yonsei University, Seoul 03722, Republic of Korea}
\address[cau]{Department of Physics, Chung-Ang University, Seoul 06974, Republic of Korea}
\address[uos]{Department of Physics, University of Seoul, Seoul 02504, Republic of Korea}
\address[anl]{Argonne National Laboratory, Lemont, IL 60439, U.S.A.}
\address[nasa]{NASA Goddard Space Flight Center, Greenbelt, MD 20771, U.S.A.}


    
\begin{abstract}
AstroPix is a high-voltage CMOS (HVCMOS) monolithic active pixel sensor (MAPS) developed for future space-based gamma-ray missions. It is also a candidate technology for the imaging layer of the Barrel Imaging Calorimeter (BIC) in the ePIC experiment at the future Electron–Ion Collider (EIC). 
We report the first AstroPix test-beam results obtained at the KEK Photon Factory Advanced Ring (PF-AR) and the CERN Proton Synchrotron (PS) T10 beam line in 2025, using the third prototype (AstroPix-v3).
AstroPix-v3 sensors were operated as both standalone tracking layers and imaging layers interleaved with prototype lead/scintillating-fiber (Pb/SciFi) calorimeter modules, using electron and hadron beams in the few-GeV/$c$ momentum range. Event synchronization between the continuous readout of AstroPix-v3 and the trigger-based readout of the Pb/SciFi calorimeter was achieved using a common timestamp. The AstroPix-v3 sensors exhibit stable performance, reaching a maximum hit efficiency of 68\% at a bias voltage of $-400$~V under pion-dominated beam conditions.
When combined with the Pb/SciFi calorimeter, the AstroPix layers successfully capture the development of electromagnetic showers. 
Using Cherenkov-based particle identification, electron-induced events exhibit significantly higher hit multiplicities and broader spatial distributions than pion-induced events, thereby providing clear discrimination between electromagnetic and hadronic showers. These results demonstrate that AstroPix-v3 provides effective, high-granularity imaging of shower development and is well-suited as an imaging layer in future calorimeter systems for both collider and space-based experiments.
\end{abstract}





\end{frontmatter}



\section{Introduction}
\label{introduction}

High-Voltage CMOS (HVCMOS)~\citep{Peric:2007zz} Monolithic Active Pixel Sensors (MAPS) are an advanced evolution of conventional CMOS MAPS. HVCMOS sensors can sustain high bias voltages to enlarge the depletion region, enabling fast charge collection and improved radiation tolerance while retaining most of the functionality of traditional CMOS MAPS. In the medium- and soft-gamma-ray energy range, particularly below approximately 10~MeV, precise two-dimensional hit-position information within a single detector layer is essential for accurate event reconstruction and effective background suppression. MAPS-based detectors provide this capability, making them an attractive technology for next-generation gamma-ray telescopes that require high spatial resolution while maintaining a low material budget.

AstroPix is an HVCMOS MAPS detector under development to address these requirements for future space-based gamma-ray missions~\citep{Steinhebel:2022ips, Caputo_2023, Suda:2023zX, Steinhebel:2025pra, Suda:2024v3per, Striebig:v4design, Suda:2026v4per}, including the All-sky Medium Energy Gamma-ray Observatory eXplorer (AMEGO-X)~\citep{Caputo:2022xpx}. The AstroPix design is derived from the ATLASPix chip, originally developed for charged-particle tracking in the ATLAS inner detector~\citep{Striebig:v4design, Peric:2019hmv}, and has since evolved through several iterations toward a sensor architecture optimized for space applications. This evolution emphasizes low electronic noise, radiation tolerance, and detector geometries tailored for gamma-ray detection, making AstroPix a promising candidate technology for upcoming medium-energy gamma-ray instruments.

Beyond its original development for space-based gamma-ray astronomy, AstroPix is also being considered as a candidate silicon sensor for the imaging layer of the Barrel Imaging Calorimeter (BIC) in the ePIC experiment at the future Electron–Ion Collider (EIC)~\citep{Klest:2024xlm, AbdulKhalek:2021gbh}. In this context, AstroPix provides fine-grained two-dimensional hit information that is crucial for precise shower imaging, particle identification, and the separation of electromagnetic and hadronic components. Its high-granularity imaging capability is particularly valuable for electron/pion ($e/\pi$) separation and for $\pi^0$ identification through detailed shower-shape reconstruction.

The current version under test, AstroPix-v3, represents the third major iteration of the AstroPix design. Compared to earlier versions~\citep{Steinhebel:2022ips, Caputo_2023, Suda:2023zX}, AstroPix-v3 is the first full-reticle implementation, featuring a $2 \times 2\mathrm~\rm{cm}^2$ sensor area with a $35 \times 35$ pixel matrix and a pixel pitch of $500 ~\unit{\um}$ \citep{Steinhebel:2025pra}. The sensor is fabricated using a standard high-voltage CMOS process with a deep n-well (DNW) structure, which forms a depletion region in the p-type bulk silicon under high-voltage bias. With an unthinned bulk thickness of 720~\unit{\um}, AstroPix-v3 is designed to provide sufficient charge collection for ionizing particles and gamma-ray–induced secondary particles.

\begin{figure*}[htbp]
    \centering
    \includegraphics[width=0.75\linewidth]{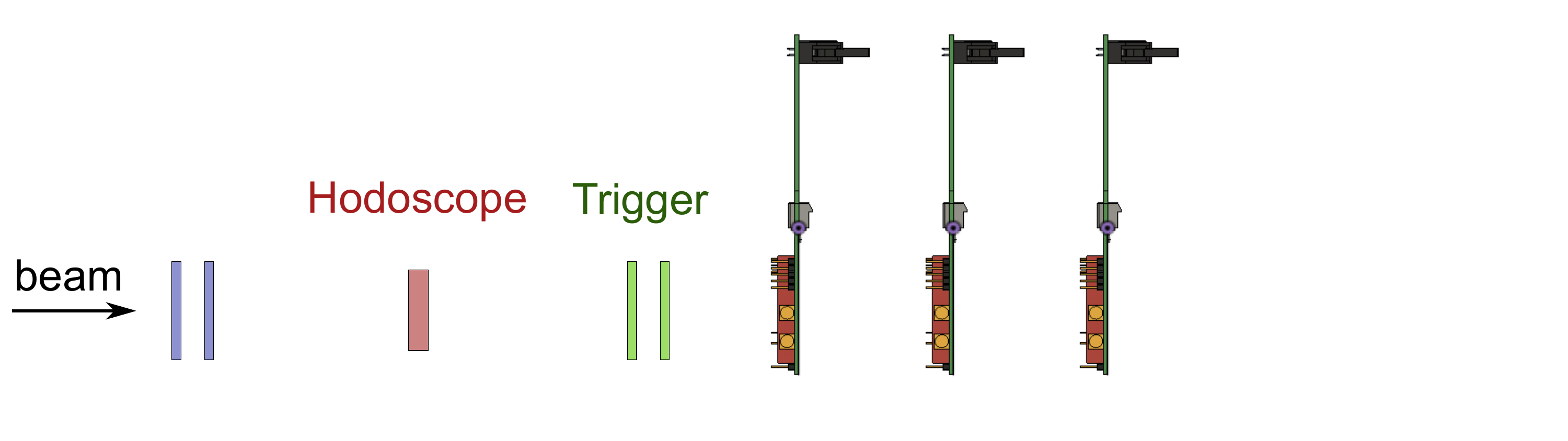}
    \vspace{0.05\linewidth}
    \includegraphics[width=0.75\linewidth]{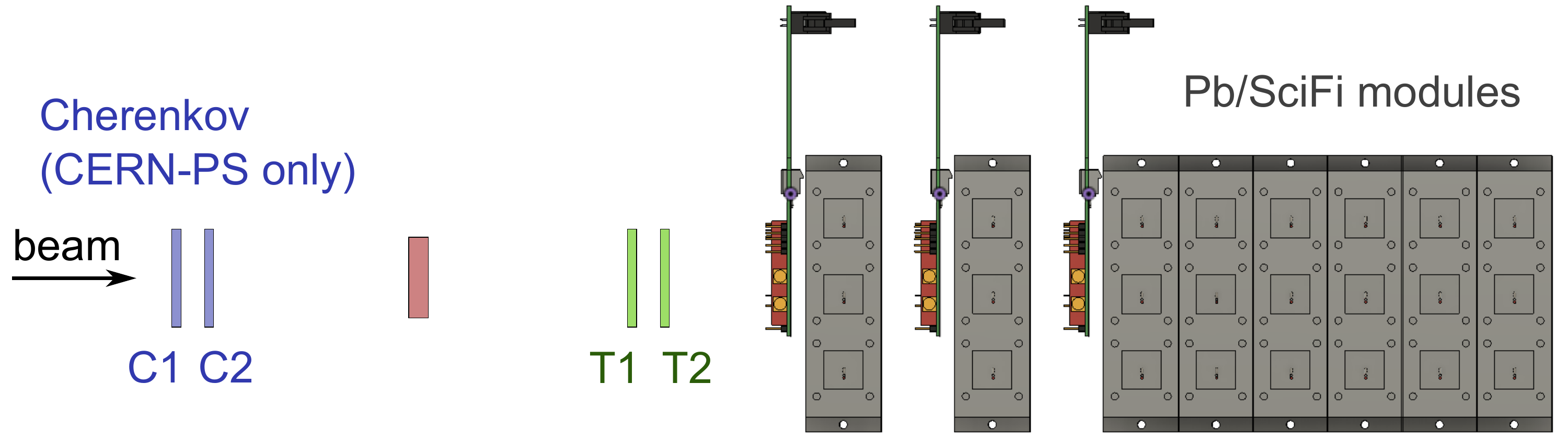}
    \caption{Schematic diagrams of the experimental setup at the CERN PS. Top: AstroPix-v3 operated in standalone mode; bottom: AstroPix-v3 interleaved with the Pb/SciFi modules.}
    \label{fig:expsetup}
\end{figure*}

The basic performance characteristics of AstroPix-v3, including noise level, energy resolution, and electrical properties, have been extensively studied by multiple institutes. However, its performance in a controlled beam environment--such as hit efficiency, position and energy resolution, timing performance, and response to electromagnetic and hadronic particles--has not yet been reported. To address this, the Korean Barrel Imaging Calorimeter (KoBIC) group conducted test-beam campaigns at two facilities in 2025: the first at the KEK Photon Factory Advanced Ring (PF-AR) test beam line~\citep{Mitsuda:2023smx} in June, followed by a second campaign at the CERN Proton Synchrotron (PS) T10 beam line~\citep{Bernhard:2021jeb} in July. The KEK PF-AR campaign, using a high-purity electron beam, focused on validating the data acquisition system and characterizing the detector response with and without the Pb/SciFi calorimeter modules.  Building on this validated setup, the CERN PS T10 campaign enabled reliable measurements of the hit efficiency. In addition, the mixed beam of electrons and hadrons, together with particle-identification detectors, enabled a direct comparison of electromagnetic and hadronic shower development.

The primary objective of these tests was to evaluate the performance of AstroPix-v3 sensors in conjunction with prototype lead/scintillating-fiber (Pb/SciFi) modules under electron and hadron beams. For this purpose, multiple AstroPix-v3 sensors mounted on single-chip carrier boards were operated in a synchronized data-acquisition scheme with the Pb/SciFi modules, enabling correlated measurements of tracking and energy deposition. In particular, by placing an AstroPix layer between Pb/SciFi modules, the experimental setup allows an investigation of the capability to discriminate between electromagnetic and hadronic showers using combined tracking and calorimeter information. This capability is an important requirement for both space-based gamma-ray instruments and the BIC at ePIC.

The remainder of this article is organized into four sections. Section 2 describes the experimental setup, including the beam conditions, trigger system, and detector layout. The data analysis procedures and selection criteria are presented in Section 3, while the test-beam results are discussed in Section 4. Finally, Section 5 provides a brief summary and discusses the significance of the results.


\section{Experimental setup}

\begin{figure*}[htbp]
    \centering
    \includegraphics[width=0.4\linewidth, angle=270]{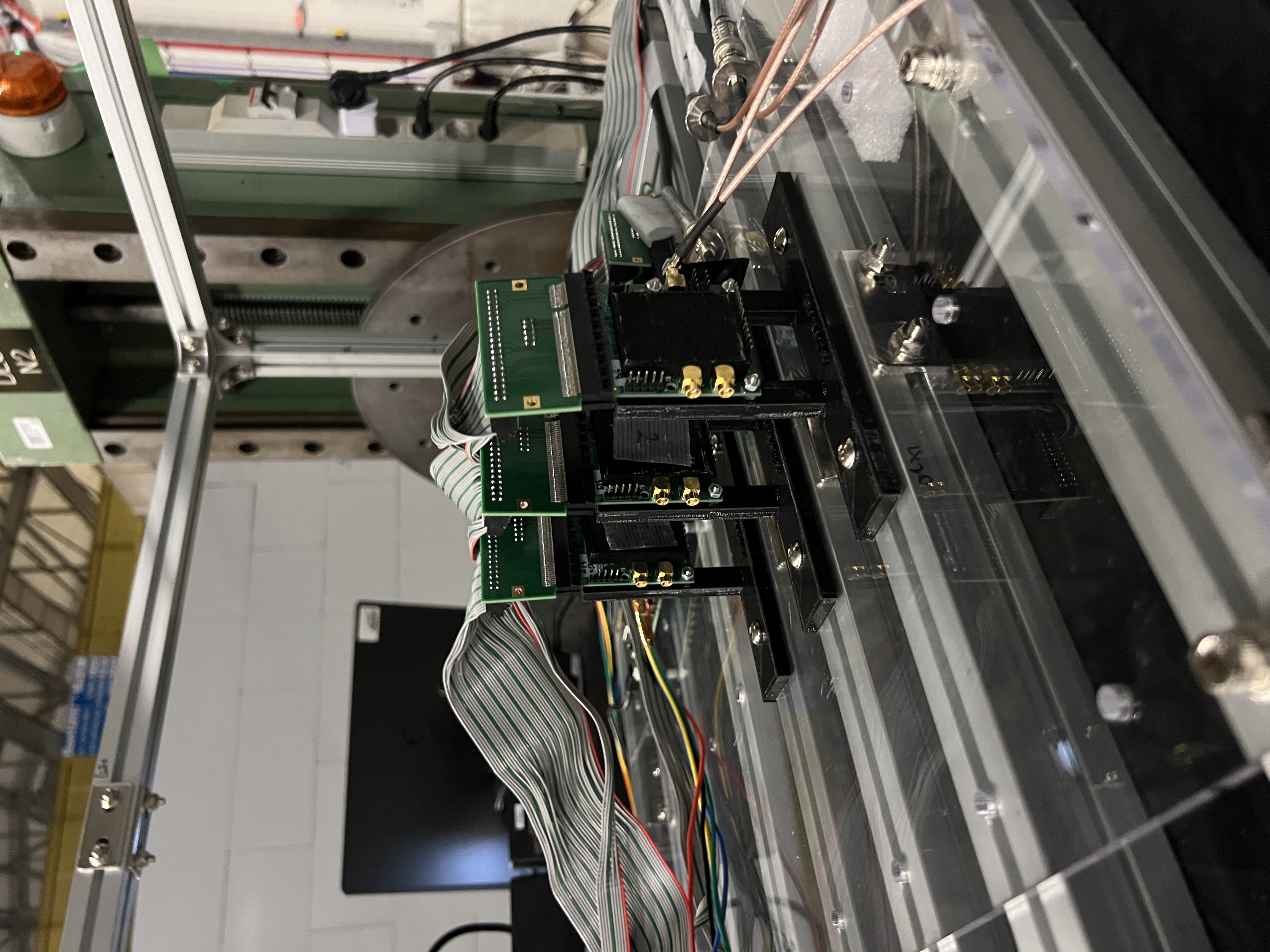}
    \hspace{0.05\textwidth}
    \includegraphics[width=0.4\linewidth, angle=270]{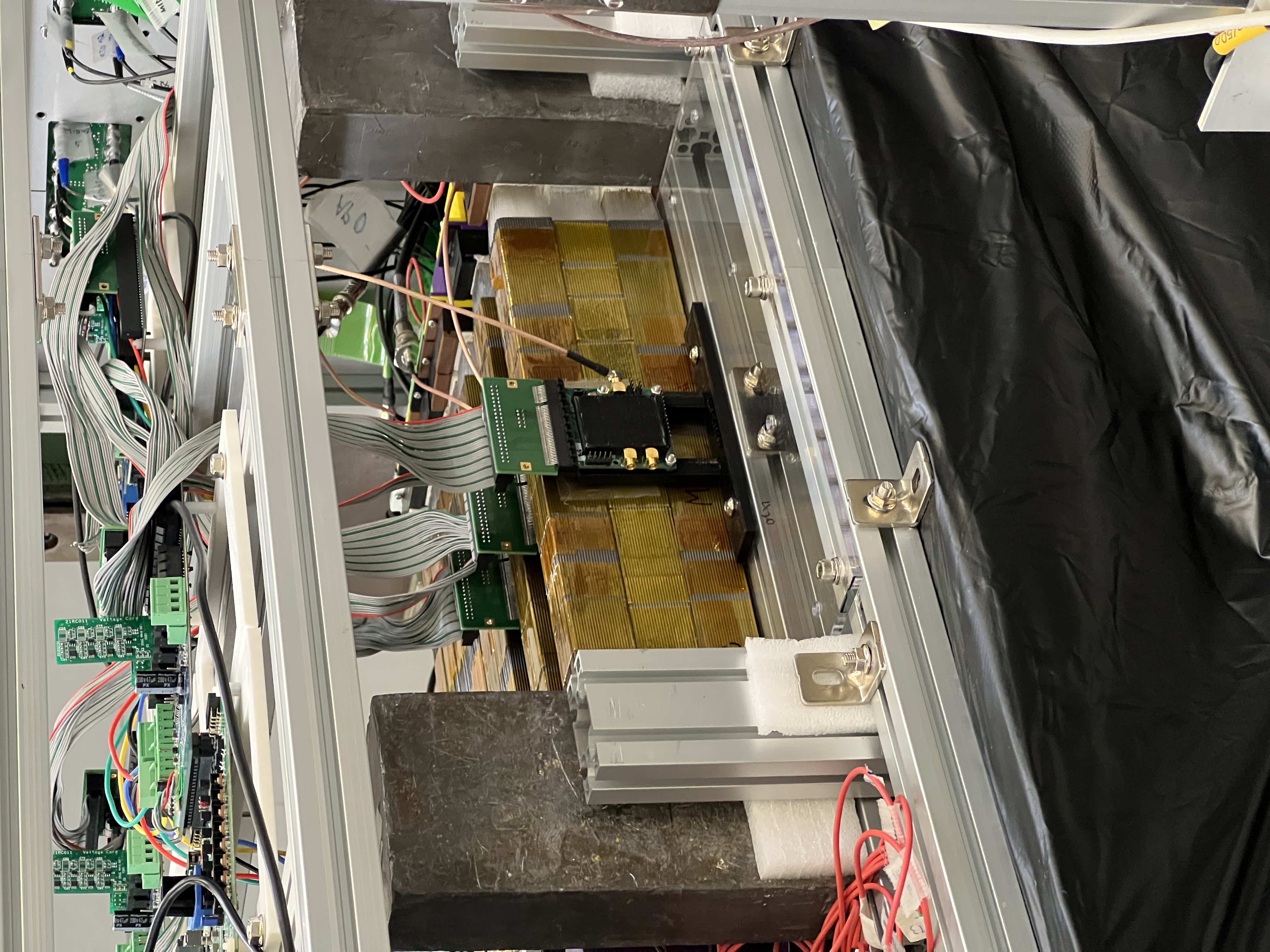}
    \caption{AstroPix-v3 operated in standalone mode (left) and interleaved with Pb/SciFi modules (right), corresponding to the top and bottom configurations in Figure~\ref{fig:expsetup}, respectively.}
    \label{fig:astropix}
\end{figure*} 

\subsection{Beam information}


The KEK PF-AR test beam line provides a high-purity electron beam for detector R\&D~\cite{Mitsuda:2023smx}. The beam originates from 6.5 or 5~GeV electrons circulating in the PF-AR storage ring.  Bremsstrahlung photons produced at a wire target are converted into electron–positron pairs, and the electrons are subsequently transported to the experimental area.  The available momentum range is typically 0.5–5~GeV/$c$. 
During the data taking, the storage ring was operated with an electron energy of 5~GeV, and electrons with a momentum of 4.5~GeV/$c$ were selected.  Except during injection into the storage ring, the beam rate remains stable over time. To suppress transient rate increases during injection, a dedicated veto signal provided by the beam line was applied. With the $15 \times 15~\mathrm{mm}^2$ trigger configuration used in this study, the typical beam rate was approximately 300~Hz.

The CERN PS T10 beam line delivers a secondary beam primarily composed of electrons, pions, and muons, produced by 24 GeV/$c$ protons impinging on a production target~\cite{Bernhard:2021jeb}. Protons and kaons are also present; however, their fractions are relatively small. The particle composition in the momentum range from 0.5 to 11.5 GeV/$c$ has been updated in a recent study~\citep{vanDijk:2025ggb}.  Two threshold Cherenkov counters installed upstream of the experimental area were used for particle identification. The gas pressure was adjusted depending on the beam momentum, and electron/pion separation was performed using the Cherenkov pulse height.
A collimator controls the beam intensity, which can be varied between approximately $10^{3}$ and $10^{7}$ particles per spill, and the momentum spread ranges from 0.6\% to 15\% depending on the selected momentum. For the present test-beam campaign, the intensity was set to approximately $10^{4}$ particles per spill. From the recorded data, the instantaneous particle rate was estimated to be about 2.7 kHz, with a spill structure corresponding to a 0.4 s flat-top.


\subsection{Detectors}

Figure \ref{fig:expsetup} illustrates the detector layout and configuration employed at the CERN PS. The setup at the KEK PF-AR was nearly identical, except for the presence of the Cherenkov detectors and minor differences in the Pb/SciFi calorimeter configuration.

The trigger detectors (T1, T2) consisted of two identical scintillator plates with dimensions of 30~$\times$~30~$\times$~2~$\mathrm{mm^3}$. The scintillation light from each plate was collected by three silicon photomultipliers (SiPMs)\footnote{S13360-3050PE from HAMAMATSU
Photonics} attached to one side of the scintillator. The trigger acceptance was defined by partially overlapping the two scintillators, forming an active area of approximately 15~$\times$~15~$\mathrm{mm^2}$.

A total of three AstroPix-v3 single-chip modules were installed along the beam direction. The spacing between adjacent chips was set to 6 cm and fixed with 3D-printed holders mounted on an acrylic plate. The acrylic plate was secured to the mechanical support frame with screws. Precision-drilled screw holes in the acrylic plate ensured accurate chip alignment, which was later verified using reconstructed beam data.
Each AstroPix-v3 chip was connected via extension cables to its corresponding GECCO readout board~\cite{Schimassek2021_1000141412}, enabling the interleaved configuration with the Pb/SciFi modules, as shown in Figure~\ref{fig:astropix}.

The prototype Pb/SciFi calorimeter consists of multiple unit modules, each with dimensions of
32~$\times$~3~$\times$~3~$\mathrm{cm^3}$. Each unit module has a material composition of Pb:SciFi:air = 40:43:17 by volume, and no adhesive was used during assembly. The effective radiation length of a single unit module is $X_{0} = 1.38$~cm. For the test at the CERN PS, a total of 3~$\times$~8 unit modules were assembled, resulting in overall dimensions of
32~$(\mathrm{h})$~$\times$~9~$(\mathrm{v})$~$\times$~24~$(\mathrm{d})$~$\mathrm{cm^3}$. 

%
%
%
The Pb/SciFi modules were arranged such that an AstroPix-v3 chip could be placed between adjacent modules, closely realizing the original design concept in which AstroPix sensors are interleaved with the calorimeter layers. The scintillation light produced in each unit module was collected by two glass photomultiplier tubes (PMTs)\footnote{R11265U-100 from HAMAMATSU Photonics} attached to both ends of the module. 
In addition, a hodoscope consisting of sixteen 1-mm-thick scintillating fibers arranged in both horizontal and vertical directions was also installed; however, its data were not used in the present analysis.


\begin{figure*}[htbp]
    \centering
    \includegraphics[width=0.65\linewidth]{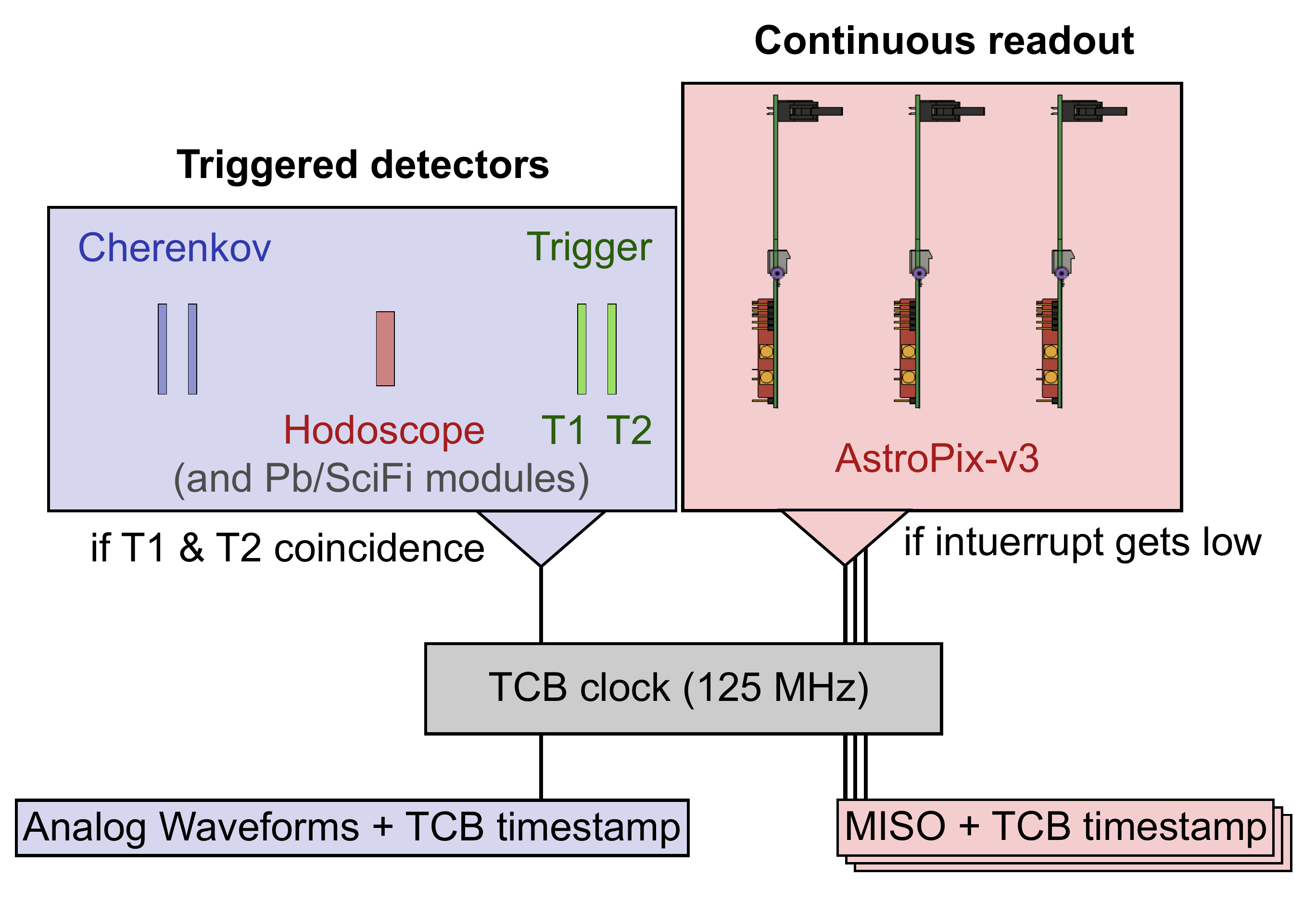}
    \caption{Schematic diagram of the DAQ configuration adopted during the test at the CERN PS. T1 and T2 denote the two trigger detectors. An interrupt low state indicates that the chip is ready to transmit data as its internal threshold condition is satisfied.
    %
    }
    
    \label{fig:daq}
\end{figure*} 

\subsection{Data acquisition}

Figure~\ref{fig:daq} illustrates the data acquisition (DAQ) configuration employed during the test at the CERN PS. 
A common global trigger was generated when the two trigger detectors (T1 and T2) coincided. This global trigger was distributed to the front-end electronics of the Pb/SciFi calorimeter, Cherenkov counters, trigger detectors, and hodoscope through a common 125 MHz trigger control board (TCB) to initiate synchronized data acquisition. The analog signals from the aforementioned detectors were digitized and recorded using flash ADC (FADC) modules.
In contrast, the AstroPix-v3 modules were operated in a continuous readout mode rather than in a trigger-based mode. To synchronize with the trigger-based readout detectors, an additional readout device was introduced to record the MISO (master-in, slave-out) signal from the SPI (serial peripheral interface) protocol. This auxiliary readout device detects hits in AstroPix-v3 by monitoring transitions in the digital interrupt signal and records the corresponding global timestamp issued by the TCB, along with the MISO data stream. Despite the fundamentally different readout schemes, global synchronization among all detectors was successfully achieved using this approach.


\section{Data analysis}

For successful event reconstruction of incident charged particles, the timestamps from the global trigger and the AstroPix-v3 must be properly matched. The trigger timestamp, $t_{\rm Trigger}$, is recorded by the TCB at the moment both trigger detectors T1 and T2 produce signals exceeding a pulse-height threshold of 100~mV, and therefore follows the actual beam passage with a delay of only a few tens of nanoseconds. In contrast, the AstroPix-v3 timestamp, $t_{\rm global}$, is the TCB timestamp latched at the falling edge of the digital interrupt signal issued by each AstroPix-v3 chip. Since this interrupt is asserted only after the on-chip time-over-threshold (ToT) evaluation is complete, $t_{\rm global}$ carries an intrinsic delay of several microseconds relative to $t_{\rm Trigger}$, necessitating an explicit timing correction.

To account for this offset, the AstroPix-v3 timestamp was corrected using the minimum ToT value ($t_{\rm min\,ToT}$) among the hit packets associated with a single interrupt (i.e., within the same readout). In multi-hit events, the digital interrupt is generated by the first signal whose positive pulse falls below the discriminator threshold, while subsequent hits are read out sequentially via the SPI interface. Since a shorter ToT corresponds to an earlier return of the pulse below the threshold, the minimum ToT provides the best estimate of the hit packet that initiated the interrupt. The corrected timestamp is defined as
\begin{equation}
    t_{\rm AstroPix} = t_{\rm global} - t_{\rm min\,ToT},
    \label{eq:timestamp}
\end{equation}
In addition, the finite SPI readout latency contributes to a spread in the observed timing. The SPI clock was operated at 8.33~MHz in a dual-SPI configuration (2 bits per clock cycle). A 10-byte hit packet, therefore, corresponds to 40 clock cycles, equivalent to approximately 4.8~$\mu$s. Since all hit information stored in the internal buffer is transferred within less than 20~$\mu$s per AstroPix-v3 readout, events were considered time-matched if the following criterion is satisfied:

\begin{equation}
    0 \leq \Delta t \leq 20~\mu{\rm s},
    \label{eq:time_match}
\end{equation}

where $\Delta t = t_{\rm AstroPix} - t_{\rm Trigger}$.
    
\begin{figure}[htb]
    \centering
    \includegraphics[width=\columnwidth]{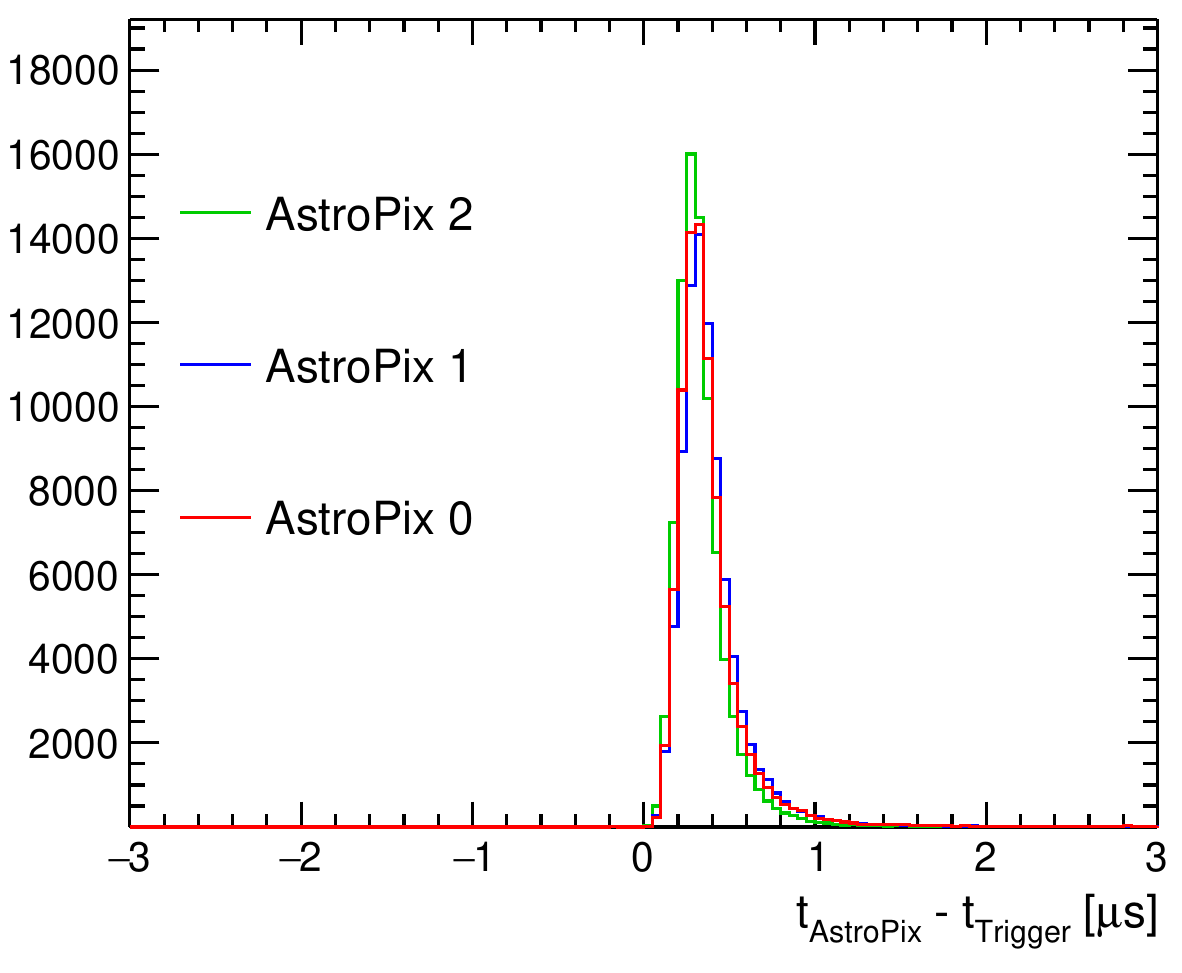}
    \caption{Measured timing offset between the global trigger and each AstroPix layer with a 4.5 GeV/c beam at KEK PF-AR. The setup did not include Pb/SciFi modules.}
    \label{fig:ana:dt}
\end{figure}

Figure~\ref{fig:ana:dt} shows the timing offset distributions between the hits recorded in the three successive AstroPix-v3 layers (AstroPix 0, 1, 2 from upstream to downstream) and the global trigger. The measurement was performed using a 4.5~GeV/$c$ electron beam at KEK PF-AR, with the AstroPix-v3 operated in standalone mode. In this configuration, each triggered event was expected to produce a single hit in each AstroPix-v3 layer.
The measured timing offsets relative to the global trigger are confined within 5~\unit{\us} for all three layers, indicating stable synchronization between the continuous readout of the AstroPix-v3 and the global trigger. The same measurement was repeated at the CERN PS, confirming the stability and reproducibility of the synchronization scheme.

\begin{figure*}[ht!]
    \centering
    \includegraphics[width=0.8\linewidth]{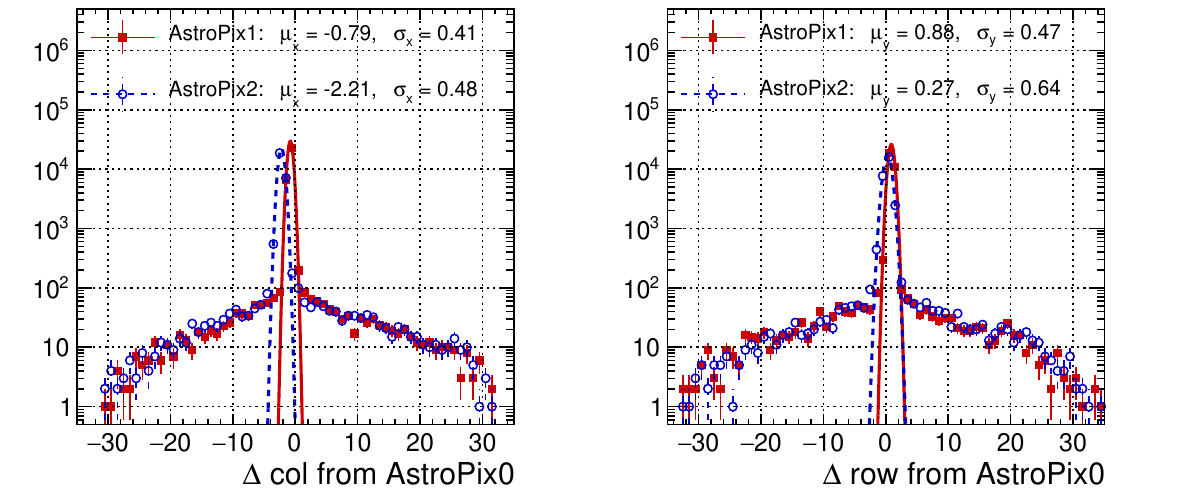}
    \caption{
    Distributions of hit-position differences in column (left) and row (right) coordinates between the first AstroPix-v3 layer and the downstream layers, measured with a 4~GeV/$c$ beam at the CERN PS. 
    Gaussian fits performed in the range $-5 < \Delta \mathrm{col (row)} < 5$ pixels are overlaid to extract the mean ($\mu$) and width ($\sigma$) values.
    }
    \label{fig:ana:correlation}
\end{figure*}

Figure~\ref{fig:ana:correlation} shows the distributions of the position differences in column ($\Delta$col) and row ($\Delta$row) coordinates between the first (beam-facing) AstroPix-v3 layer and the downstream layers, obtained with a 4~GeV/$c$ beam at the CERN PS, with an approximate composition of 16\% electrons and 84\% pions.  For this analysis, only events containing exactly one hit in each layer were selected to ensure unambiguous track association.
The narrow Gaussian structures centered near zero demonstrate the validity of the event-matching procedure based on the common TCB timestamp.  The strong concentration of events within a small $\Delta$ region indicates that hits assigned to the same trigger correspond to the same traversing particle, confirming reliable synchronization between the continuous-readout AstroPix system and the triggered beam instrumentation.
Small but non-zero offsets are observed in the mean values of the distributions. In particular, the extracted $\mu_x$ values for the second and third layers are $-0.79$ and $-2.21$ pixels, respectively. These offsets are attributed to a small relative misalignment between the AstroPix layers. Possible sources include a slight tilt of the supporting acrylic plate or non-uniform mechanical mounting of the AstroPix modules. To compensate for this effect, subsequent analyses apply a position correction defined as
\begin{equation}
    \centering
    x_i = \mathrm{col}_i - \mu_{x,i}, \qquad
    y_i = \mathrm{row}_i - \mu_{y,i} \quad (i=1, 2)
\end{equation}
where $\mu_{x,i}$ and $\mu_{y,i}$ are the mean offsets obtained from the Gaussian fits in Figure~\ref{fig:ana:correlation}. Although the hit coordinates originate from discrete pixel indices, subtracting the fitted mean offsets enables sub-pixel alignment of the detector layers.

The fitted widths are narrow in all cases, demonstrating stable detector operation and consistent track reconstruction. Events outside the Gaussian core are predominantly associated with noisy pixels. Therefore, for track reconstruction in the standalone configuration, a conservative selection requirement of $|\Delta \mathrm{x}| < 2$ and $|\Delta \mathrm{y}| < 2$ was applied to suppress noise contributions. This validated track definition provides a robust basis for the efficiency measurement presented in the following section, where the outer layers are used to define reference tracks to evaluate the intrinsic hit-detection probability of the middle layer.


\section{Results}

\subsection{Hit efficiency of AstroPix-v3}

The hit efficiency of the AstroPix-v3 was evaluated in standalone mode without Pb/SciFi modules. The measurement was performed using an 8~GeV/$c$ beam at the CERN PS. At this momentum, the beam is dominated by pions, thereby reducing the impact of multiple Coulomb scattering, which can degrade the accuracy of the efficiency determination.

To ensure that only events corresponding to beam particles traversing the AstroPix-v3 layers were considered, the efficiency of the second layer was measured using the first and third layers as reference planes. Events were selected in which synchronized hits were present in both the first and third layers and coincided with the global trigger, thereby defining a track candidate. To suppress noise hits and pile-up effects, only events with exactly one hit in each reference layer were selected. A spatial-correlation requirement was then applied: The reconstructed hit positions in all three layers were required to be mutually consistent within $|\Delta x| < 2$ and $|\Delta y| < 2$ pixels.

\begin{figure}[ht!]
    \centering
    \includegraphics[width=0.9\columnwidth]{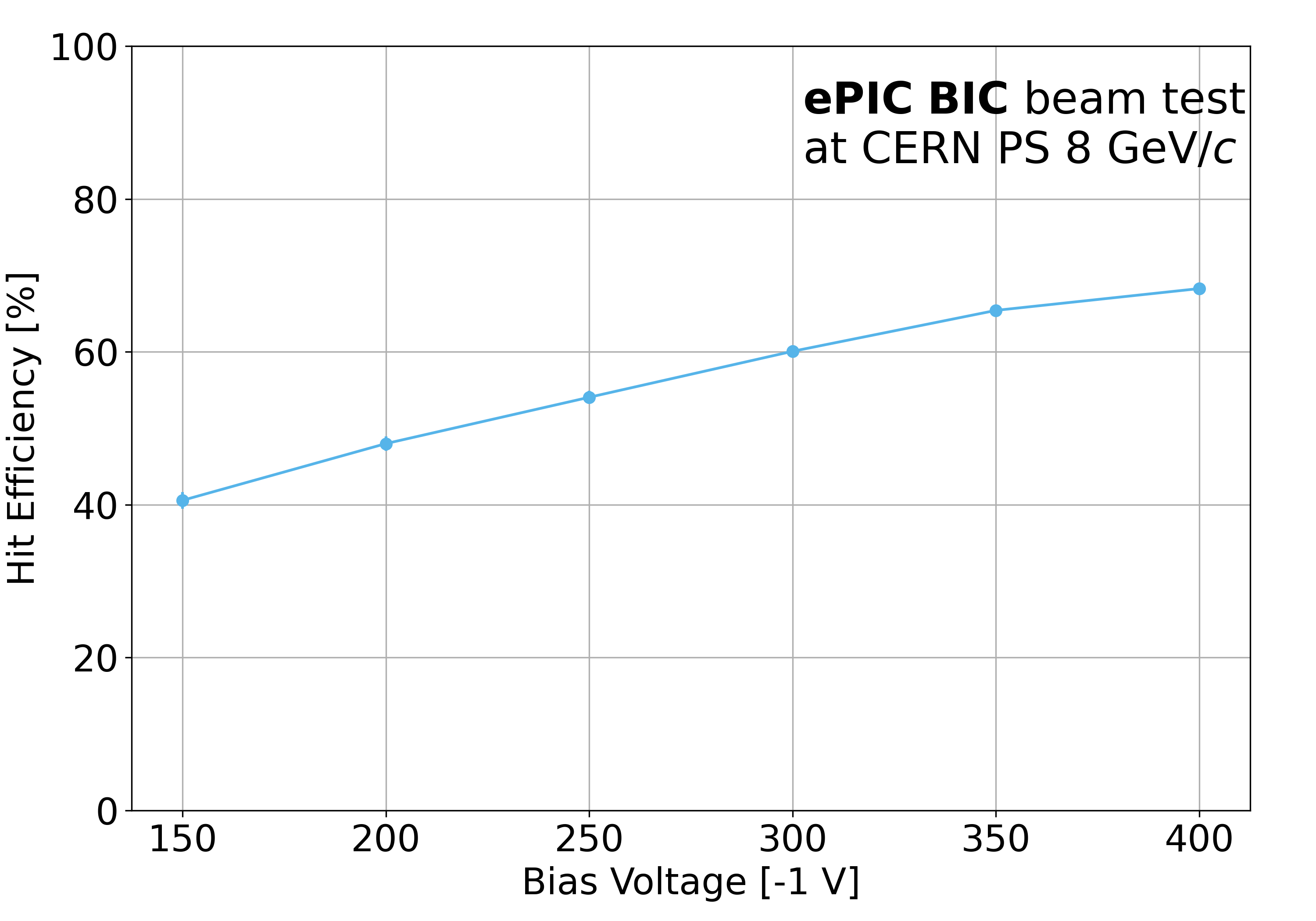}
    \caption{Hit efficiency of an AstroPix-v3 layer measured with an 8 GeV/$c$ beam at the CERN PS as a function of bias voltage ($-150$ to $-400$ V). Efficiency is defined as the ratio of hits in the target layer to those in the reference layers. All hits were required to satisfy a spatial-correlation requirement of $|\Delta x| < 2$ and $|\Delta y| < 2$ pixels.}
    \label{fig:res:efficiency}
\end{figure} 

The discriminator threshold of the AstroPix-v3 was fixed at 200~mV for all measurements. The hit efficiency was measured as a function of the applied bias voltage, which was varied from $-150$ V to $-400$ V. As shown in Figure~\ref{fig:res:efficiency}, the efficiency increases monotonically with increasing bias voltage, reaching approximately 68\% at $-400$ V, compared to about 40\% at lower bias voltages.

The increase in efficiency is attributed to the increased depletion depth at higher bias voltages, which enhances charge collection from traversing charged particles. For AstroPix-v3, the depletion depth is estimated to be approximately 100~\unit{\um} ~\citep{Suda:2024v3per}. This limited depletion depth is consistent with the relatively modest maximum efficiency observed at the highest applied bias voltage. Future iterations of the AstroPix sensor, such as AstroPix-v5, are expected to achieve near-full depletion up to approximately 500~\unit{\um}, which should substantially improve hit efficiency.


\subsection{Measurement of electromagnetic shower development}
\label{sec:emshower}

\begin{figure*}[ht!]
    \centering
    \includegraphics[width=0.8\linewidth]{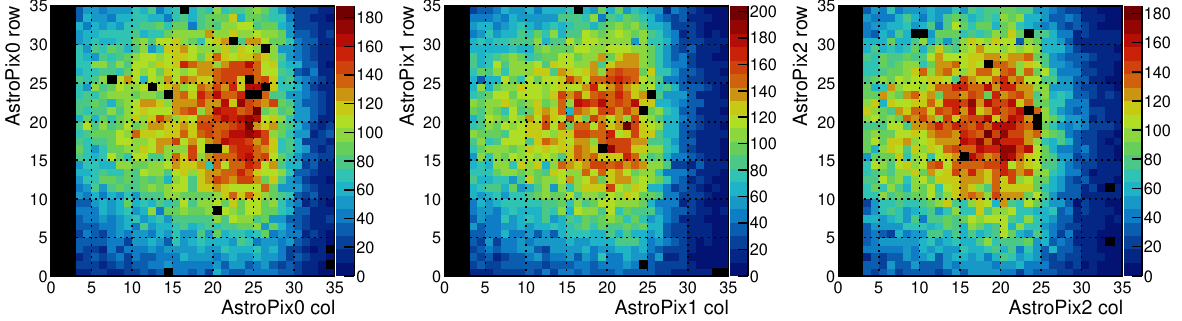}
    \includegraphics[width=0.8\linewidth]{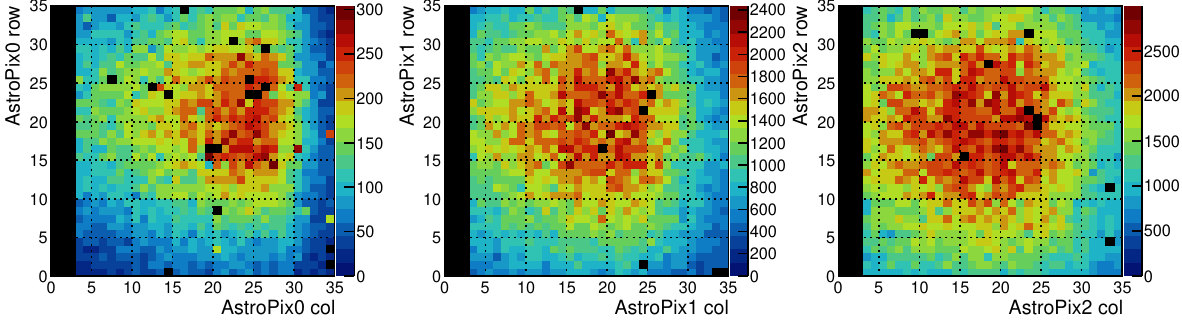}
    \caption{Hit maps of the AstroPix-v3 layers measured using a 4.5 GeV/$c$ electron beam at KEK PF-AR. Top: standalone configuration; bottom: configuration interleaved with Pb/SciFi modules. The black squares indicate masked-out noisy pixels.
    }
    \label{fig:res:hitmap}
\end{figure*}

\begin{figure}[ht!]
    \centering
    \includegraphics[width=\columnwidth]{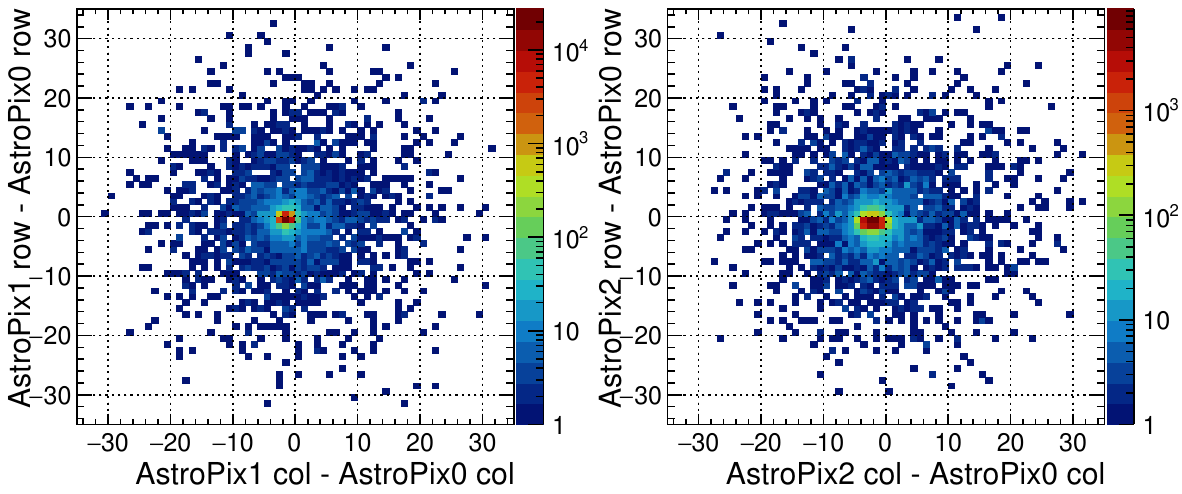}
    \includegraphics[width=\columnwidth]{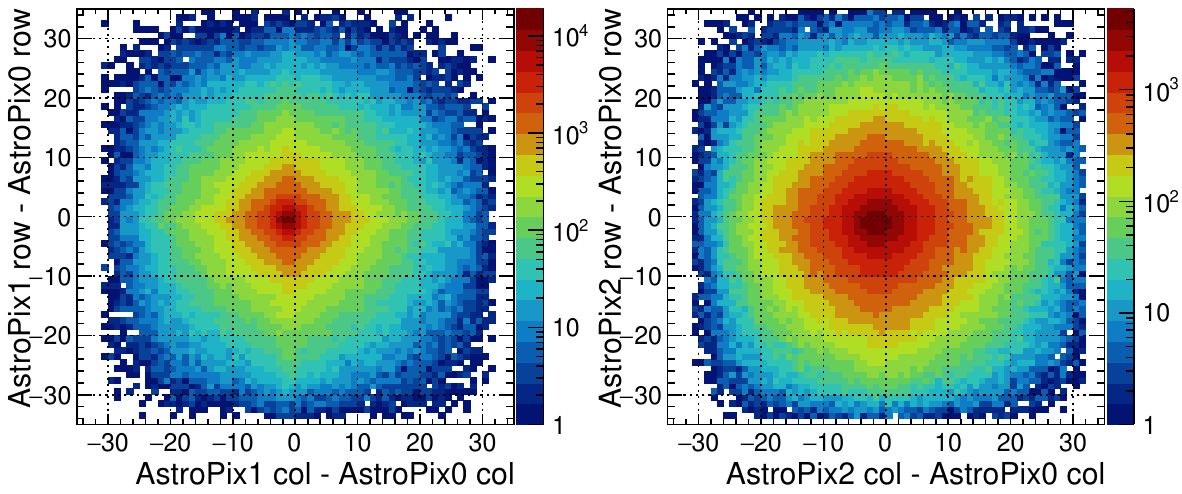}
    \caption{Position residuals between the first (reference) AstroPix-v3 layer and the other layers measured using a 4.5 GeV/c electron beam at KEK PF-AR. Top: standalone; bottom: interleaved with Pb/SciFi modules.}
    \label{fig:res:posdiff}
\end{figure}

Figure~\ref{fig:res:hitmap} shows the hit maps of the three AstroPix-v3 layers measured using a 4.5~GeV/$c$ electron beam at KEK PF-AR. In standalone mode, the hit maps of all three layers exhibit similar spatial distributions, consistent with single charged particles traversing the detector with minimal secondary-particle production.
In contrast, when AstroPix-v3 is interleaved with Pb/SciFi modules, the hit distributions progressively broaden from upstream to downstream layers. This behavior is consistent with the development of an electromagnetic (EM) shower, in which secondary electrons and photons are generated within the Pb/SciFi volume and produce an increasing lateral spread as they propagate downstream.

Figure~\ref{fig:res:posdiff} shows the position residuals of the second and third AstroPix layers with respect to the first (upstream) layer, providing a more quantitative characterization of the EM shower development. Compared with the standalone configuration, the interleaved configuration exhibits a significantly broader radial distribution of hits. Furthermore, within the interleaved configuration, the third layer shows a wider radial distribution than the second layer, indicating continued lateral development of the EM shower.



\subsection{Separation between the EM and hadronic showers}
\label{sec:showersep}

As briefly discussed in Section~1, one key requirement for AstroPix to be adopted as an imaging sensor for the BIC is the ability to distinguish EM showers from hadronic showers. The beam at the CERN PS T10 beam line is well suited for this study, as it contains both electrons and hadrons, predominantly pions.
The beam line provides particle identification (PID) through measurements of the Cherenkov-counter pulse height. For beam momenta of 4 GeV/$c$ and above (below), events with Cherenkov pulse heights exceeding 30 (100) ADC counts were classified as electrons. This PID capability enables the selection of EM- or hadron-dominated event samples, allowing a direct evaluation of the EM–hadronic shower separation performance of the combined AstroPix–Pb/SciFi system, which forms the basic concept of the BIC.

The shower-separation performance was tested using the AstroPix-v3 configuration interleaved with Pb/SciFi modules. In this configuration, the first AstroPix layer was located in the upstream of the first Pb/SciFi module and served as a reference plane for the incoming particle and the other two AstroPix-v3 layers. The second and third AstroPix-v3 layers were located downstream of the first and second 3~cm-thick Pb/SciFi modules to probe shower development downstream of them.

\begin{figure}[ht!]
    \centering
    \includegraphics[width=\columnwidth]{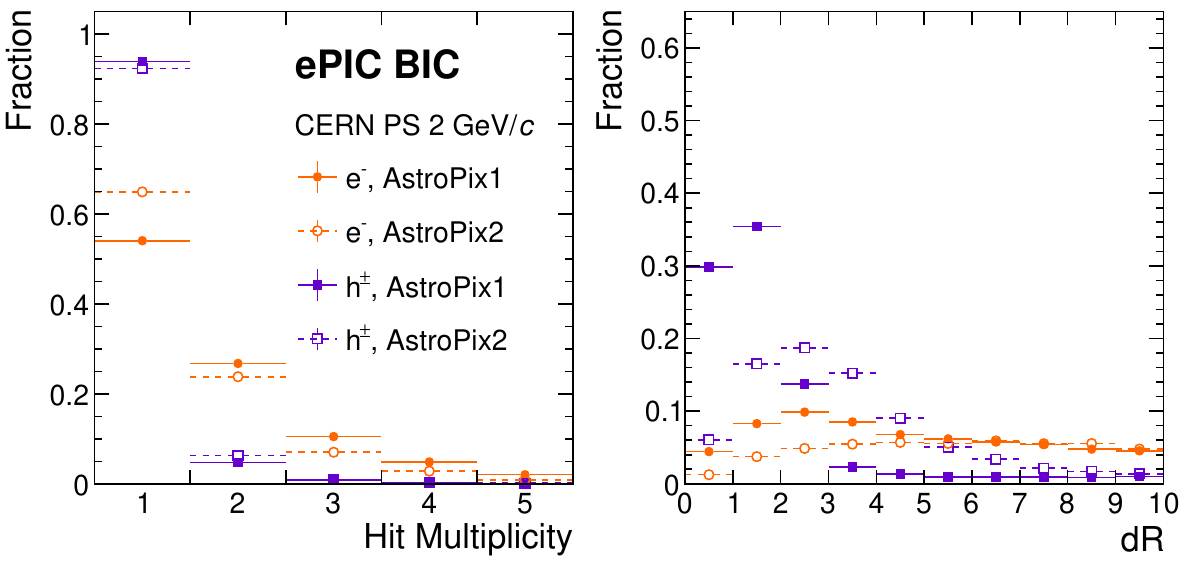}
    \includegraphics[width=\columnwidth]{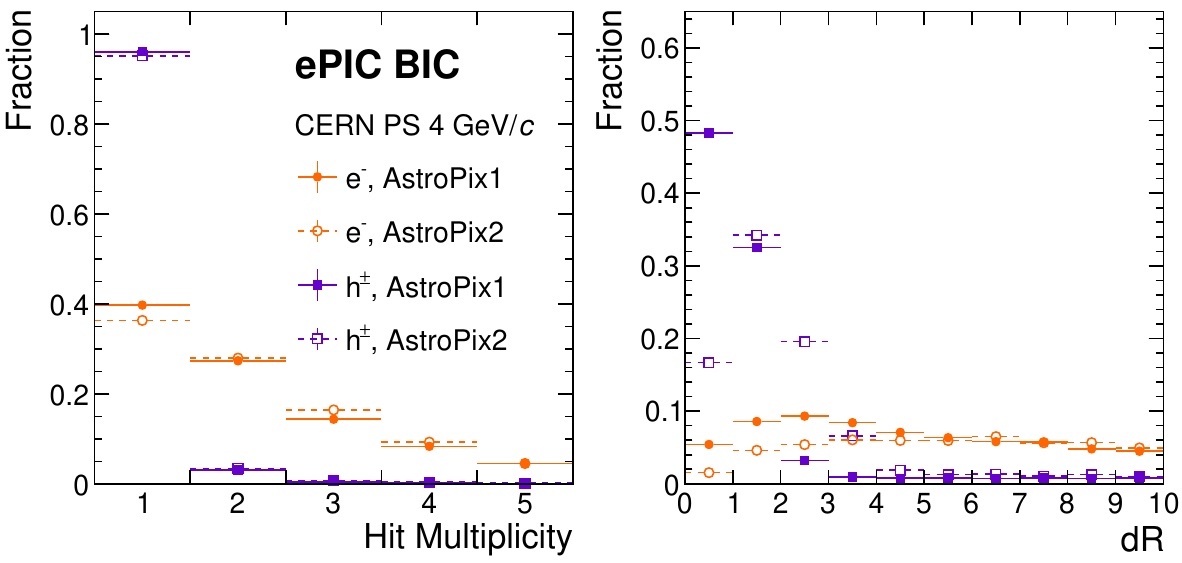}
    \caption{Hit multiplicity (left) and radial position residuals with respect to the first layer (right) for the second (AstroPix1) and third (AstroPix2) AstroPix-v3 layers. Top panels correspond to 2~GeV/$c$ beam data, and bottom panels to 4~GeV/$c$ beam data. The radial residual is defined as $dR = \sqrt{(\Delta x)^2 + (\Delta y)^2}$ using the corrected hit positions. All distributions are normalized to the total number of particle-identified events (electrons or hadrons), such that the vertical axis represents the fraction of events per bin.}
    \label{fig:res:ehshower}
\end{figure} 

Figure~\ref{fig:res:ehshower} presents the hit multiplicity distributions in the second and third AstroPix-v3 layers, together with the radial position residuals relative to the first layer. The radial position residual was defined as $dR = \sqrt{(\Delta x)^2 + (\Delta y)^2}$, where $\Delta x$ and $\Delta y$ denote the differences in the corrected hit coordinates defined in Equation (3). The distributions are normalized to the total number of particle-identified events (electrons or hadrons), such that the vertical axis represents the fraction of events in each bin. To suppress pileup effects, only events with a single hit in the first layer were selected. Furthermore, at least one hit was required in either the second or third layer to exclude events with missing downstream hits due to particles scattered outside the detector acceptance or detector inefficiencies.


The hit multiplicity distributions exhibit a clear separation between electron- and pion-induced events. For pions, the distributions are dominated by single-hit events at both 2 and 4 GeV/$c$, with no significant dependence on beam momentum. This behavior indicates that the current Pb/SciFi configuration is not sufficiently thick for pions to develop distinguishable hadronic showers based solely on hit multiplicity. In contrast, the multiplicity of electron-induced events increases markedly with beam momentum. The fraction of events with two or more hits reaches approximately 40\% at 2 GeV/$c$ and exceeds 60\% at 4 GeV/$c$.

A similar trend is observed in the radial position residuals relative to the first layer. Electron-induced events exhibit significantly broader residual distributions than pion-induced events, consistent with the lateral development of electromagnetic showers in the Pb/SciFi modules. In contrast, pion-induced events remain narrowly distributed, reflecting the delayed onset and limited development of hadronic showers within the calorimeter volume.


For electron-induced events, the lateral spread shows little dependence on beam momentum within the studied range, suggesting that the transverse shower size is similar for different beam energies. It is also observed that the $dR$ values in the third layer are generally larger than those in the second layer, reflecting the continued lateral development of the electromagnetic shower as it propagates downstream. For hadron-induced events, the residual distributions remain narrower than those of electrons, but they still broaden toward the downstream layers due to multiple scattering. In addition, the effect of multiple scattering becomes smaller with increasing beam energy, resulting in a narrower $dR$ distribution at higher momenta.

Based on the results presented in Sections \ref{sec:emshower} and \ref{sec:showersep}, the AstroPix-v3 sensor interleaved with Pb/SciFi modules demonstrates the capability to distinguish EM from hadronic showers in various momenta, using hit multiplicity and spatial correlation observables.

\section{Summary}

This paper presents the first test-beam results of the AstroPix-v3 sensor operated both in standalone mode and interleaved with Pb/SciFi modules. The measurements were performed using an electron beam at KEK PF-AR and a pion-dominated hadron beam at the CERN PS T10 facility, covering a momentum range of a few GeV/$c$. The AstroPix-v3 demonstrated stable operation under beam conditions. A robust synchronization scheme based on a common TCB clock enabled reliable event matching between the continuous readout of AstroPix-v3 and the trigger-based readout of the other detectors, as confirmed by consistent timing offsets and clear spatial correlations among the AstroPix-v3 layers.

The hit efficiency of AstroPix-v3 was measured using an 8 GeV/$c$ pion-dominated beam at the CERN PS. An efficiency of approximately 68\% was achieved at a bias voltage of $-$400 V, consistent with the partial depletion depth of the current sensor design. The observed dependence on bias voltage is consistent with expectations and suggests that future iterations with increased depletion depth are likely to achieve substantially higher efficiencies.
When interleaved with Pb/SciFi modules, the AstroPix-v3 layers successfully resolved the development of EM showers. The hit maps and position-residual distributions exhibit a clear broadening of the spatial hit pattern relative to the upstream reference layer, demonstrating AstroPix-v3's capability to function as a high-granularity imaging layer in a sampling calorimeter. These results support the suitability of AstroPix sensors for detailed shower-imaging applications.
A direct comparison between EM and hadronic showers was performed at the CERN PS using Cherenkov-based PID. Compared with pion-induced events, electron-induced events exhibit higher hit multiplicity and significantly broader radial-position residual distributions in the layers downstream of the Pb/SciFi modules. The fraction of multi-hit events in the downstream layers is comparable at 2 and 4 GeV/$c$, indicating that EM shower development is already substantial at the lower energy for the given Pb/SciFi thickness. In contrast, higher energies primarily influence the shower's lateral extent. The observed separation demonstrates the effectiveness of the AstroPix–Pb/SciFi system in distinguishing EM from hadronic showers using imaging observables.

In summary, these results indicate that AstroPix is a viable imaging sensor for calorimetric applications. The demonstrated performance supports its prospective use as an imaging layer in the BIC of the ePIC experiment at the future EIC, as well as its continued development for space-based medium-energy gamma-ray missions. 

\section*{Acknowledgements}
This work was supported by the National Research Foundation of Korea (NRF) grants funded by the Korean government (MSIT) (Grant No. RS-2024-00408265, 2020R1A2C3013540, RS-2025-00514606, RS-2024-00355188, RS-2021-NR062024, RS-2023-00280845, RS-2018-NR031074, RS-2022-NR070380, and RS-2023-00279977).
This work also received funding from the European Union's Horizon Europe Research and Innovation Programme under Grant Agreement No. 101057511 (EURO-LABS). 
The material is based upon work partially supported by the U.S. Department of Energy, Office of Science, Office of Nuclear Physics and Laboratory Directed Research and Development funding from Argonne National Laboratory, provided by the Director, Office of Science, of the U.S. Department of Energy under Contract No. DE-AC02-06CH11357.
The authors would like to thank the CERN PS team and the KEK PF-AR team for their support.

\appendix



\bibliographystyle{elsarticle-num} 
\bibliography{references}

\end{document}